\documentclass[english,aps,prb,twocolumn,nofootinbib,showkeys,secnumarabic,superscriptaddress,floatfix]{revtex4}

\usepackage[english]{babel}

\usepackage{amssymb,amsmath} 

\usepackage[pdftex]{graphicx}
\usepackage[caption=false]{subfig}
\usepackage{color} 
\usepackage[usenames,dvipsnames]{xcolor}
\usepackage[colorlinks=true,linkcolor=RoyalBlue,citecolor=OliveGreen,urlcolor=OliveGreen,linktoc=page]{hyperref} 

\usepackage{enumerate}
\usepackage{grffile}
\usepackage{balance}
\usepackage{float}

\newcommand{\sys}{\mathrm{\scriptscriptstyle S}}
\newcommand{\res}{\mathrm{\scriptscriptstyle R}}
\newcommand{\bL}{\langle\mathrm{L}|}
\newcommand{\bR}{\langle\mathrm{R}|}
\newcommand{\cL}{|\mathrm{L}\rangle}
\newcommand{\cR}{|\mathrm{R}\rangle}
\newcommand{\LL}{\mathrm{\scriptscriptstyle LL}}
\newcommand{\RR}{\mathrm{\scriptscriptstyle RR}}
\newcommand{\LR}{\mathrm{\scriptscriptstyle LR}}
\newcommand{\RL}{\mathrm{\scriptscriptstyle RL}}
\newcommand{\I}{{\hat 1}}
\newcommand{\B}{{\scriptscriptstyle\rm B}}
\newcommand{\C}{{\scriptscriptstyle\rm C}}
\newcommand{\E}{{\scriptscriptstyle E}}
\newcommand{\tr}{\mathrm{tr}}
\newcommand{\Su}{{\hat S}_{\uparrow}}
\newcommand{\Sd}{{\hat S}_{\downarrow}}

\begin{document}

\title{$H$-Theorem in Quantum Physics}

\author{G. B. Lesovik}
\affiliation{L.D. Landau Institute for Theoretical Physics RAS, Akad. Semenova av., 1-A, Chernogolovka, 142432, Moscow Region, Russia}
\affiliation{Theoretische Physik, Wolfgang-Pauli-Strasse 27, ETH Z\"urich, CH-8093 Z\"urich, Switzerland}

\author{A. V. Lebedev}
\affiliation{Theoretische Physik, Wolfgang-Pauli-Strasse 27, ETH Z\"urich, CH-8093 Z\"urich, Switzerland}

\author{I. A. Sadovskyy} 
\affiliation{Materials Science Division, Argonne National Laboratory, 9700 S. Cass Av., Argonne, Illinois 60637, USA}

\author{M. V. Suslov} 
\affiliation{Moscow Institute of Physics and Technology, Institutskii per. 9, Dolgoprudny, 141700, Moscow District, Russia}

\author{V. M. Vinokur} 
\affiliation{Materials Science Division, Argonne National Laboratory, 9700 S. Cass Av., Argonne, Illinois 60637, USA}

\date{September 12, 2016}

\begin{abstract}
Remarkable progress of quantum information theory (QIT) allowed to formulate mathematical theorems for conditions that data-transmitting or data-processing occurs with a non-negative entropy gain. However, relation of these results formulated in terms of entropy gain in quantum channels to temporal evolution of real physical systems is not thoroughly understood. Here we build on the mathematical formalism provided by QIT to formulate the quantum $H$-theorem in terms of physical observables. We discuss the manifestation of the second law of thermodynamics in quantum physics and uncover special situations where the second law can be violated. We further demonstrate that the typical evolution of energy-isolated quantum systems occurs with non-diminishing entropy.
\end{abstract}


\maketitle

\section{Introduction}

In the 1870-s, Ludwig Boltzmann published his celebrated kinetic equation and the $H$-theorem\cite{Boltzmann:1872,Boltzmann:1896} that gave the statistical foundation of the second law of thermodynamics.\cite{Lebowitz:1999} The $H$-theorem states that if $f(x, v, \tau)$ is the distribution density of molecules of the ideal gas at the time $\tau$, position $x$ and velocity $v$, which satisfies the kinetic equation, then entropy defined as $S = -\int dx \, dv \, f(x, v, \tau)\log f(x, v, \tau)$ is non-diminishing, i.e. that $dS/d\tau \geqslant 0$. Boltzmann's kinetic equation rests on the molecular chaos hypothesis which assumes that velocities of colliding particles are uncorrelated and independent of position. Striving to bypass molecular chaos hypothesis, unjustified within the classical mechanics, John von Neumann proposed\cite{Neumann:1929} pure quantum mechanical origin of the entropy growth. He defined entropy through quantum mechanical density matrix ${\hat\rho}$ as $S(\hat{\rho}) = -k_\B \tr \{ \hat{\rho} \log\hat{\rho} \}$, and offered a proof of non-decreasing entropy resting on the final procedure of macroscopic measurement. As this proof yet invoked concepts going beyond pure quantum mechanical treatment, the nonstop tireless search for the quantum mechanical foundation of the $H$-theorem have been continuing ever since, see Ref.~\onlinecite{Gemmer:2009} for a review. At the same time there have been a remarkable progress in quantum information theory (QIT), which formulated several rigorous mathematical theorems about the conditions for a non-negative entropy gain.\cite{Nielsen:2011,Holevo:2012} In this communication we show how the results of QIT apply to physical quantum systems and phenomena establishing thus non-diminishing von Neumann's entropy in physics and formulate the conditions under which the evolution accompanied by non-diminishing entropy arises within pure quantum mechanical framework.

To describe quantum dynamics of an open system, the quantum information theory introduces the so-called \textit{quantum channel} (QC) defined as a trace-preserving completely positive map, ${\Phi(\hat\rho)}$, of a density matrix.\cite{Nielsen:2011} A remarkable general result of the QIT states that the entropy gain in a channel ${\Phi(\hat\rho)}$ is\cite{Holevo:2010}
\begin{equation}
	S(\Phi(\hat\rho)) - S(\hat\rho)
	\geqslant -k_\B \tr
	\{\Phi({\hat\rho})\log\Phi(\I) \},
	\label{eq:entropy_gain}
\end{equation}
where $\I$ is the identity operator. This formula was derived from the monotonicity property\cite{Lindblad:1975} of the relative entropy under the quantum channel $\Phi$: $S(\Phi(\hat\rho)||\Phi(\hat\sigma)) \leqslant S(\hat\rho || \hat\sigma)$, where $S(\hat\rho || \hat\sigma) = \tr \{ \hat\rho(\log(\hat\rho)-\log(\hat\sigma)) \}$. There exists a wide class of channels, the so-called \textit{unital} channels, defined by the relation $\Phi(\I) = \I$, for which the right hand side of Eq.~\eqref{eq:entropy_gain} vanishes, $\tr \{\Phi({\hat\rho}) \log\Phi(\I) \} = 0$, so that the entropy gain is non-negative, $\Delta S \equiv S({\Phi(\hat\rho)}) - S({\hat\rho}) \geqslant 0$. Then within the framework of the QIT one can formulate the quantum $H$-theorem as follows: the entropy gain during evolution is nonnegative if the system evolution can be described by the unital channel. Moreover, for a quantum system endowed with the finite $N$-dimensional Hilbert space, the unitality condition becomes not only a sufficient, but also the necessary condition for non-diminishing entropy. Indeed, let us assume that for any initial state of a system with $N$-dimensional Hilbert space, the entropy gain in a channel $\Phi$ is non-negative. It then follows that for the chaotic state ${\hat\rho_\C} = \I/N$ that already has the maximal entropy, $S({\hat{\rho}}_\C) = k_\B \ln N$, the entropy cannot grow, $S(\Phi({\hat\rho}_\C)) = S({\hat\rho}_\C)$. Thus $\Phi({\hat\rho}_\C)={\hat\rho}_\C$, therefore, $\Phi(\I) = \I$ and the channel is unital. For an infinite-dimensional quantum system the entropy is not continuous,\cite{Wehrl:1978} and this situation requires special consideration. Finally, it is noteworthy that there exist certain classes of states that evolve with $\Delta S \geqslant 0$ even if the channel is not unital.\cite{Amosov:2015}

To connect the general result (\ref{eq:entropy_gain}) and the related mathematical $H$-theorem formulation to the realm of physics note that any quantum system interacting with the reservoir generates a quantum channel. Indeed, let us consider joint evolution of the grand system, comprising a given quantum system and a reservoir initially prepared in a disentangled state, $\mathcal{\hat P} = \hat\rho \otimes \hat\rho_\res$, where $\hat\rho_\res$ is the density matrix of the reservoir. Let ${\hat U}$ be the unitary operator describing the temporal evolution of the grand system. Then, according to the Stinespring-Kraus dilation theorem\cite{Nielsen:2011} $\Phi(\hat\rho) = \tr_\res \{ {\hat U} \mathcal{\hat P} {\hat U}^\dag \}$ is the quantum channel. Note that the evolution from an initially entangled state may be accompanied by the arbitrary gain in entropy. This demonstrates the necessity of the disentanglement condition. Examples of how the initially entangled system can evolve with the decreasing entropy are given e.g. in Refs.~\onlinecite{Argentieri:2014,Lesovik:2013}.

To compare how do the classical and quantum $H$-theorems work, note that for the former to hold, the classical distribution function of the system involved should obey the kinetic equation. The constraint imposed on the evolution of the density matrix of the quantum system is that the corresponding quantum channel is unital. This defines our task as to find out the necessary conditions under which the temporal dynamics of a quantum system endowed with the specific interaction with environment can be modeled by the unital quantum channel. Below, we formulate these conditions for the so called quasi-isolated quantum systems with the negligible energy exchange with environment and demonstrate how do they apply to generic exemplary physical realizations.

\section{Quantum $H$-theorem}

In physics the positive entropy gain, according to the second law of thermodynamics, is ensured by the energy isolation of the evolving system. In contrast to the classical formulation of the second law where any isolated classical system evolves with the non-diminishing entropy, its literal extension onto the quantum case is meaningless since the entropy of any isolated quantum system does not change, $S({\hat U}(t) \, \hat\rho \, {\hat U}^\dag(t)) = S(\hat\rho)$. Hence, to bring the thermodynamic meaning to the consideration of quantum evolution one has to allow an interaction with the environment and establish the notion of the \textit{quasi-isolated system}. However, letting an arbitrarily system-environment interaction causes immediate problem. The energy exchange $\Delta E$ between the system and environment at temperature $T$ and its entropy gain are related in classical thermodynamics as $\Delta S = \Delta E /k_\B T$. One would expect that in a quantum case the similar relation also might hold provided the quantum system interacted with the macroscopic reservoir during the sufficiently long time. Moreover, long time evolution of a quantum system which exchanges energy with an environment can not, in general, be described by the unital channel at all. Indeed, consider a finite dimensional quantum system with~$N$ discrete non-degenerate energy states $|E_n\rangle$ initially prepared in the chaotic state $\hat\rho_\C = (1/N) \sum_n |E_n\rangle \langle E_n|$. Then the long-time interaction with the low temperature environment drives the system into the low energy states and hence the resulting quantum channel becomes non-unital, $\Phi(\hat\rho_\C) \neq \hat\rho_\C$. Therefore, one has to restrict allowable interactions to the class of interactions that provide the system's entanglement with the environment, but yet keep the energy exchange with the environment negligible. Such an interaction, for example, is realized for a specific environment of nuclear spins which possesses a highly energy-degenerate ground state. For a general situation of the environment endowed with the low-energy excitations, one can employ the concept of the quasi-isolated system provided there is a time separation between the dephasing time $T_2$ of the off-diagonal elements of the density matrix and the relaxation time of its diagonal elements, $T_1$. Then in the intermediate time evolution regime $T_2 < \tau \ll T_1$ the system gets entangled with its environment but its energy exchange remains still negligible. Accordingly, in what follows we discuss the systems energy-isolated from the reservoir. Furthermore, we will be assuming that our systems are initially disentangled from the reservoir.

Let us consider a fixed energy subspace $E$ of the system Hilbert space spanned by the orthonormal basis states $|\psi_{i,\E}\rangle$, $\hat{H}_\sys |\psi_{i,\E}\rangle = E|\psi_{i,\E}\rangle$, where index $i$ denotes all the remaining non-energy system's degrees of freedom and $\hat{H}_\sys$ is the system Hamiltonian. It is convenient to present the evolution operator, ${\hat U}$, of the grand system (a system plus reservoir) as
\begin{equation}
	{\hat U} 
	= \sum_{\E,ij} |\psi_{j,\E}\rangle \langle \psi_{i,\E}| \, 
	s_{ji,\E} {\hat F}_{ji,\E},
	\label{eq:QES}
\end{equation}
where $s_{ji,\E}$ are the components of the scattering matrix corresponding to the transition amplitude between the system's quantum states $|\psi_{i,\E}\rangle \to |\psi_{j,\E}\rangle$ (without taking into account interaction with the reservoir) and operators ${\hat F}_{ji,\E}$ are the family of operators acting in the reservoir Hilbert space, with the subscripts $i$, $j$, and $E$ specifying the system's states (for details see Appendix~\ref{sec:unitarity_constraints}). The factorization into $s_{ji,\E}$ and ${\hat F}_{ji,\E}$ is not unique, so we will be choosing the most suitable one for each particular case.

For an energy-isolated quantum system the quantum states at different energies transform independently. To determine whether the evolution belongs in the class of the unitality channel, one has to check if the system obeys the $\Phi(\I) = \I$ relation. Using the unitarity of ${\hat U}$, one finds
\begin{equation}
	\Phi_{jj'}(\I_\E)
	- [\I_\E]_{jj'}
	= \sum_i s_{ji,\E} s_{j'i,\E}^* \,
	\langle [{\hat F}_{j'i}^\dag, {\hat F}_{ji}] \rangle,
	\label{eq:PHII}
\end{equation}
where $\langle \ldots \rangle$ is averaging with respect to the initial state of the reservoir, and $\I_\E=\sum_{i}|\psi_{i,\E}\rangle\langle\psi_{i,\E}|$ (the proof is presented in Appendix~\ref{sec:theorem_proof}). This relation is our central result. It establishes the criterion for unitality of the energy-isolated system in terms of physical operators describing the interaction of the quantum system with the reservoir.
Combining the concept of unitality and relation (\ref{eq:PHII}) we reformulate quantum $H$-theorem as follows.

\textit{Let the quantum system interacting with the reservoir be initially disentangled from it and be energy-isolated during the evolution. Let $\sum_i s_{ji,\E} s_{j'i,\E}^* \, \langle [{\hat F}_{j'i}^\dag, {\hat F}_{ji}] \rangle=0$, where operators ${\hat F}_{ji,\E}$ and coefficients $s_{ji,\E}$ are defined as in Eq.~\eqref{eq:QES}, for energies $E$ at which the system can be found with a finite probability, i.e. $\langle\psi_{\E,i}|\hat\rho|\psi_{\E,i}\rangle>0$. Then the resulting quantum channel is unital in the subspace spanned by the states $|\psi_{\E,i}\rangle$ with a finite $\langle\psi_{\E,i} |\hat\rho|\psi_{\E,i}\rangle>0$ and hence the quantum system evolves with a non-negative entropy gain $\Delta S = S(\Phi(\hat\rho)) - S(\hat\rho) \geqslant 0$.}

\begin{figure}[t]
	\begin{center}
		\includegraphics[width=8.0cm]{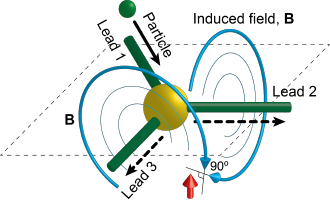}
	\end{center} \vspace{-4mm}
	\caption{
		Scattering in a 3-lead setup. A particle incident from the lead 1 is scattered 
		into two other leads 2 and 3. Propagating particle induces magnetic field 
		perpendicular to the lead direction. The spin is placed at the point where the respective 
		fields induced by particles propagating along leads 2 and 3 are perpendicular 
		to each other. To simplify consideration, we choose the set up design allowing to neglect 
		the field induced by the particle in the lead 1.
	}
	\label{fig:Y-junction}
\end{figure}

There may be two major scenaria by which the right hand side of the Eq.~\eqref{eq:PHII} can vanish: (i)~The `microscopic' scenario where the reservoir operators $\hat{F}_{ji}$ commute individually, $[{\hat F}_{j'i}^\dag, {\hat F}_{ji}] = 0$; and (ii)~The `macroscopic' scenario where only the averaged commutators vanish, $\langle [{\hat F}_{j'i}^\dag, {\hat F}_{ji}]\rangle = 0$, while individual operators do not. Below we demonstrate that the scenario (i) is realized for the electron interacting with the phonon bath, under the condition of the quasi-elastic scattering, see Eq.~\eqref{eq:Fphonon}. Here the unitality of the quantum channel appears already on a microscopic level for every electron-phonon collision event. The scenario (ii) realizes, for example, for an electron interacting with the random ensemble of three-dimensional (3D) nuclear spins, so that the vanishing of the averaged commutators occurred at the macro-level in the thermodynamic limit of the large spin ensemble. Importantly, the above formulation of the $H$-theorem applies also to a situation where vanishing of the weighted commutators $\sum_i s_{ji,\E} s_{j'i,\E}^* \, \langle [{\hat F}_{j'i}^\dag, {\hat F}_{ji}] \rangle$ occurs only within certain energy range and does not hold for the arbitrary energies of the system. For example, in case of the electron-phonon interaction the dynamics of an electron can be described by the unital channel only at high electron energies exceeding the Debye energy, see below.

The obtained formulation of the quantum $H$-theorem enables us to reveal a fundamental difference in how the second law of thermodynamics manifests itself in quantum and classical physics. In classical thermodynamics the energy-isolated system inevitably evolves with non-diminishing entropy. We find that in quantum physics the situation is different. To demonstrate that, we construct an energy-isolated quantum system for which $\langle [{\hat F}_{j'i}^\dag, {\hat F}_{ji}] \rangle \neq 0$ and which thus evolves with the negative entropy gain. Let us consider a charged particle moving in a three-lead conductor and interacting with the spin via the induced magnetic field, see Fig.~\ref{fig:Y-junction}, and, according to our general framework, initially disentangled from the spin. In the absence of the external magnetic field the energy exchange is absent and the particle is energy-isolated. The joint scattering states of the particle and spin have the form
\begin{equation}
	|\psi_\alpha^{\scriptscriptstyle(\mathrm{in})}\rangle |\sigma_0\rangle 
	\to \sum_\beta s_{\beta\alpha}\, |\psi_\beta^{\scriptscriptstyle(\mathrm{out})}\rangle \, 
	{\hat U}_\beta {\hat U}_\alpha^\dag \, |\sigma_0\rangle,
	\label{eq:fork_state}
\end{equation}
where $|\psi_\alpha^{\scriptscriptstyle(\mathrm{in/out})}\rangle$ is the particle's incoming/outgoing state in the lead $\alpha$, $s_{\beta\alpha}$ are the components of the scattering matrix of the three lead set up, $|\sigma_0\rangle$ is the initial state of the spin and ${\hat U}_\alpha$ (${\hat U}_\alpha^\dag$) is the unitary spin-$1/2$ rotation of the spin due to outgoing (incoming) electron in the lead $\alpha$. Then the operators ${\hat F}$ of Eq.~\eqref{eq:PHII} are defined as ${\hat F}_{\beta\alpha}={\hat U}_\beta {\hat U}_\alpha^\dag$.

Let us recall now that rotations of a spin about different axis in general do not commute. We choose spin-$1/2$ rotations as ${\hat U}_1 = \I$, ${\hat U}_2 = i\hat\sigma_x$ and ${\hat U}_3 = i\hat\sigma_y$, where $\hat\sigma_x$ and $\hat\sigma_y$ are the Pauli matrices, so that $[{\hat U}_\alpha, {\hat U}_\beta] \neq 0$. Accordingly, $\langle [{\hat F}_{\beta' \alpha}^\dag, {\hat F}_{\beta\alpha}] \rangle\neq 0$, and the resulting quantum channel is non-unital. The explicit calculation gives (for details of derivation see Appendix~\ref{sec:non_unital_channel})
\begin{align}
	\Phi(\I_\E)
	= \I_\E + 2i \, \bigl\{ 
	& |1\rangle \langle 2|\, s_{13,\E} \, s_{23,\E}^*\, \langle\hat\sigma_x\rangle
	\nonumber \\
	& + |1\rangle \langle 3| \,s_{12,\E} \, s_{23,\E}^*\, \langle\hat\sigma_y\rangle
	\nonumber \\
	& - |2\rangle \langle3| \, s_{12,\E} \, s_{13,\E}^*\, \langle\hat\sigma_z\rangle
	- \mathrm{H.c.} \bigr\}.
	\label{eq:PHII3}
\end{align}
Let the initial state of the spin be a pure state $|S_0\rangle = (|{\uparrow}_x\rangle + |{\uparrow}_y\rangle + |{\uparrow}_z\rangle)/{\sqrt{3}}$, so that all the $\langle\hat\sigma_\alpha\rangle$ in the Eq.~\eqref{eq:PHII3} are equal to $1/3$. Hence all off-diagonal elements of $\Phi(\I_\E)$ appear finite. Taking all $s = 2/3$, at some energy $E_0$, we construct the normalized initial state of the particle as ${\hat\rho} = (1/6) \int dE\,f(E-E_0)\I_\E$, with $f(E)$ being the normalized to unity distribution function centered around $E = 0$ and rapidly decaying as $|E|\to\infty$, and obtain $\Delta S \approx -0.05\,k_\B$. We thus demonstrated that even the energy isolation does not guarantee the evolution with the non-diminishing entropy. Note that in the discussed example the reservoir acts as some quantum analogue of the classical Maxwell demon. Namely, having been prepared in a special state, the reservoir is able to decrease the entropy of the system without the energy exchange with it, and can be referred to as a `quantum Maxwell demon' discussed in Ref.~\onlinecite{Lloyd:1997} in the context of the work extraction in nano-devices. An extension of the Second Law, accounting for the classical correlation between the system and an information reservoir, i.e. classical Maxwell demon, has recently been considered in Refs.~\onlinecite{Defner:2013,Horowitz:2014}. In what was discussed above, an electron interaction with the quantum spin does not induce any correlations between the electron and the spin and, therefore, no classical correlations are present. Hence an important distinction between how do quantum and classical Maxwell's demons operate.

\section{Elastic scattering}

\begin{figure}[b]
	\begin{center}
		\subfloat{\includegraphics[width=8.2cm]{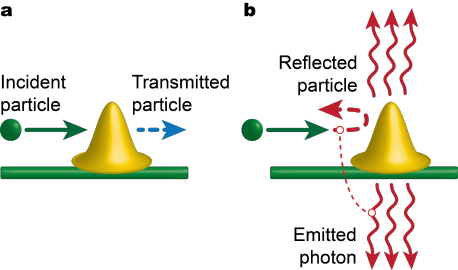} \label{fig:braking_radiation_forward}}
		\subfloat{\label{fig:braking_radiation_backward}}
	\end{center} \vspace{-4mm}
	\caption{
		Braking radiation in 1D.
		(a)~A scattering where electron is transmitted without photon emission.
		(b)~A backward scattering event accompanied by the emission of photons.
	}
	\label{fig:braking_radiation}
\end{figure}

As a first example of a system satisfying quantum $H$-theorem, we consider an electron elastically scattered by the one-dimensional (1D) potential barrier, see Fig.~\ref{fig:braking_radiation}. Let the electron reflection be accompanied by the change of the reservoir state (e.g. let the reflection to imply emission of the low energy photon via braking radiation, or scattering of the photon), see Fig.~\subref{fig:braking_radiation_backward}, and the electron transmission to retain the state of the reservoir, see Fig.~\subref{fig:braking_radiation_forward}. Accordingly, the joint scattering states of the particle with the fixed energy assume the form,
\begin{align}
	& \cL |n\rangle
	\to s_\RL \cR |n\rangle
	+ s_\LL \cL |n'\rangle,
	\label{eq:scattering_state1} \\
	& \cR |n\rangle
	\to s_\LR \cL |n\rangle
	+ s_\RR \cR |n''\rangle,
	\label{eq:scattering_state2}
\end{align}
where indices $\{\mathrm{L}, \mathrm{R}\}$ denote the incident (scattered) states in the left and right leads, respectively, $|n\rangle$ is the initial reservoir state, and $|n'\rangle$ and $|n''\rangle$ are the reservoir states resulting from the backscattering to the left and to the right, respectively. The unitality of the corresponding channel follows now from general Eq.~\eqref{eq:PHII}. However, for illustrative purpose we derive unitality straightforwardly using the explicit form of Eq.~\eqref{eq:scattering_state1} and Eq.~\eqref{eq:scattering_state2} (for details see Appendix~\ref{sec:1d_scattering}). Namely, calculating $\I_\E$ in the basis of left and right scattering channels, $i,j \in \{\mathrm{L}, \mathrm{R}\}$, we obtain
\begin{align}
	\I_\E \to \,
	& \cR \bR \, (|s_\RR|^2 + |s_\RL|^2) +
	\cL \bL \, (|s_\LR|^2+|s_\LL|^2) \, +
	\nonumber \\
	& \Bigl\{\cR \bL \, (s_\LR^*s_\RR^{\phantom *} \langle n''|n\rangle +
		s_\LL^* s_\RL^{\phantom *} \langle n|n''\rangle) + \mathrm{H.c.} \Bigr\}
	\label{eq:1Didentity}
\end{align}
and taking into account the unitarity of the overall transformation, we arrive at $\Phi(\I_\E) = \I_\E$. Since this condition holds for any $E$, then for any state the system evolves with $\Delta S \geqslant 0$. The above consideration with some minor modifications holds for the grand system where the role of the reservoir is taken up by a single spin located near the scatterer. Then the spin remains intact if the particle is reflected and is rotated by the magnetic field induced by the transmitted particle.

Let us generalize the above consideration onto the particle propagating along the two-dimensional array of scatterers and spins (comprising the reservoir) located in the $xy$-plane. The magnetic field induced by the propagating particle is perpendicular to the plane, and all spins experience the commuting unitary rotations around the perpendicular $z$-axis. All the rotations commute, hence the condition of the quantum $H$-theorem is satisfied and $\Delta S \geqslant 0$.

Remarkably, the property of unitality of the grand system with the spin reservoir preserves in the 3D case. As we mentioned above, the rotations experienced by an individual spin may appear, in general, non-commuting. Note, however, that operators ${\hat F}$ in Eq.~\eqref{eq:PHII} acquire the form
\begin{equation}
	{\hat F}_{ji} =
	{\hat U}_{ji}^{\scriptscriptstyle (1)} \otimes \ldots \otimes {\hat U}_{ji}^{\scriptscriptstyle(N)},
\end{equation}
where ${\hat U}_{ji}^{(a)}$ is a unitary rotation of a spin $a$ by the electron experienced a scattering process from the state $|i\rangle$ to the state $|j\rangle$. Then
\begin{equation}
	[{\hat F}_{j'i}^\dag, {\hat F}_{ji}] = \prod_{a=1}^N \otimes \;{\hat U}_{j'i}^{\dag{\scriptscriptstyle(a)}} {\hat U}_{ji}^{\scriptscriptstyle(a)} - \prod_{a=1}^N \otimes\; {\hat U}_{ji}^{\scriptscriptstyle(1)} {\hat U}_{j'i}^{\dag{\scriptscriptstyle(1)}},
	\label{eq:commutator}
\end{equation}
where $N\gg 1$ is the total number of spins.
For most of the spins, the factors that appear upon averaging Eq.~\eqref{eq:commutator} are small, $\langle {\hat U}_{j'i}^{\dag{\scriptscriptstyle(a)}} {\hat U}_{ji}^{\scriptscriptstyle(a)} \rangle,\,\langle {\hat U}_{j i}^{\scriptscriptstyle(a)} {\hat U}_{j'i}^{\dag{\scriptscriptstyle(a)}} \rangle
<q<1$. We arrive at the estimate
\begin{equation}
	\langle [{\hat F}_{j'i}^\dag, {\hat F}_{ji}] \rangle\lesssim\exp[-N\ln(1/q)],
\end{equation} 
therefore, $\langle [{\hat F}_{j'i}^\dag, {\hat F}_{ji}] \rangle \to 0$ as $N \to \infty$. We see that in the macroscopic limit of the number of spins, the averaged commutators appearing in the condition for the $H$-theorem vanish in spite of the fact that the commutators for the individual spins could remain finite. Thus the evolution of the considered 3D system occurs with $\Delta S \geqslant 0$.

To proceed further, we note, that if one can find a basis in the reservoir Hilbert space where all ${\hat F}_{ji}$-operators are diagonal, then the operators ${\hat F}_{ji}$ commute. Below we present two generic physical examples where this basis can be found explicitly: (i)~an electron interacting with adiabatic two-level impurities and (ii)~electron-phonon interaction in solids.

\begin{figure}[t]
	\begin{center}
		\includegraphics[width=8.0cm]{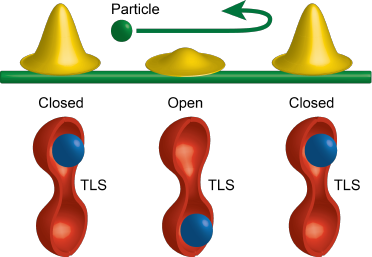}
	\end{center} \vspace{-4mm}
	\caption{
		One-dimensional random walk of an electron. Two-level systems (TLSs)
		shown as double well potentials are located equidistantly along the wire.
		Each TLS forms an effective potential for the electron, which depends
		on the TLS's quantum state. For simplicity we consider a completely
		transparent (open) or completely reflective (closed) effective scattering
		potential depending on the TLS state. At each scattering event the set
		of TLSs is replaced by a new (unentangled) one.
	}
	\label{fig:qubits}
\end{figure}

\section{Scattering on two-level systems}

Consider the electron scattering on impurities that fluctuate between two positions with nearly equal energies. To be concrete we focus on the random walk of the electron along the ensemble of TLS (see the description of the similar systems in Refs.~\onlinecite{Aharonov:1993,Childs:2002}) as shown in Fig.~\ref{fig:qubits}. Assume for simplicity that (i)~the dynamics of a TLS is slow hence its state does not change during the interaction with the electron and that (ii)~each TLS which is in the state $|{\uparrow}\rangle$ ($|{\downarrow}\rangle$), scatters the electron via elastic unitary matrix $\Su$ ($\Sd$). Then the global unitary transformation is given by ${\hat U} = \prod_n \otimes {\hat U}_n$ where ${\hat U}_n = \Su ^{\scriptscriptstyle(n)} |{\uparrow}_n\rangle \langle{\uparrow}_n| + \Sd^{\scriptscriptstyle(n)} |{\downarrow}_n\rangle \langle{\downarrow}_n|$ describes the scattering on the $n$-th impurity. Denote the scattering state of the electron moving in a direction $\mathbf{k}$ by $|\mathbf{k}\rangle$. Then the reservoir operator ${\hat F}_{\mathbf{k}\mathbf{k}'} = \prod_n\otimes {\hat F}_{\mathbf{k}\mathbf{k}'}^{\scriptscriptstyle(n)}$ with ${\hat F}_{\mathbf{k}\mathbf{k}'}^{\scriptscriptstyle(n)} = [\Su^{\scriptscriptstyle(n)}]_{\mathbf{k} \mathbf{k}'} |{\uparrow}_n\rangle \langle{\uparrow}_n| + [\Sd^{\scriptscriptstyle(n)}]_{\mathbf{k} \mathbf{k}'} |{\downarrow}_n\rangle \langle{\downarrow}_n|$, see Eq.~\eqref{eq:QES}, commute with each other and, therefore, the second term in Eq.~\eqref{eq:PHII} vanishes. Hence, in each particular scattering event the entropy of the particle is non-diminishing.

The question that now arises is whether the non-diminishing entropy maintains for the sequence of scatterings. Recall that in order for the entropy to grow monotonously, an electron should be disentangled from the TLS with which it is going to interact. Since in the course of the evolution an electron may return to the TLS on which it have scattered and, with which, therefore, it could have been getting entangled in the past, these returns would violate this `initial non-entanglement condition.' Thus to guarantee the evolution with non-diminishing entropy the TLS should have interacted with some other degrees of freedom that would lead to the memory loss of this TLS prior to the possible return of the electron. This memory loss is the manifestation of the so-called \textit{monogamy} of entanglement\cite{Coffman:2000} which is a specific property of the entanglement distribution between quantum systems: if a TLS is already entangled with an electron and later it becomes entangled with another degree of freedom then the initial entanglement with the electron vanishes. Thus the process of sequential scattering of an electron satisfies the $H$-theorem if the typical entanglement time for the TLS is less than the typical return time of an electron to the particular TLS.

\begin{figure}[b]
	\begin{center}
		\includegraphics[width=8.5cm]{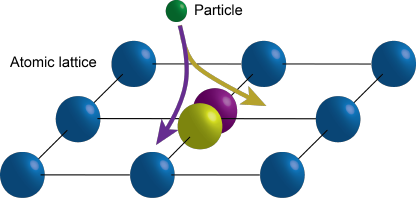}
	\end{center} \vspace{-4mm}
	\caption{
		Electrons and phonons in an atomic lattice. The scattering
		state of an electron depends strongly on the position
		of the scattering atom in the lattice. Importantly,
		the position of the scatterer remains almost unchanged
		during the scattering process because of the significant
		mass difference between the scattered electron and
		the scattering atom.
	}
	\label{fig:lattice}
\end{figure}

The interaction of a quantum system with a memoryless environment can be described by Markovian (or equivalently Lindblad) master equation.\cite{Lindblad:1976} The system's dynamics governed by Lindblad master equation can be described within the so called \textit{collision model}\cite{Ziman:2002} where a quantum system (electron) interacts locally in time with the different environmental degrees of freedom or sub-environments. In a situation, where different sub-environments are initially uncorrelated and the system interacts with the given sub-environment at most once, the resulting quantum channel possesses a \textit{divisibility} property: $\Phi = \Phi_N \circ \dots \circ \Phi_1$, where $\Phi_i$ is a quantum channel corresponding to the scattering on $i$th TLS followed by a free unitary evolution. In a more realistic situation, the TLSs may retain the partial entanglement with the electron that induces finite time memory effects in the environment. In this situation while the quantum channel is not divisible and hence cannot be described by the Lindblad master equation any more, it still can be described within in the collision model framework.\cite{Rybar:2012,Ciccarello:2013,Caruso:2014} This may result, in general, in a non-monotonic entropy evolution. This issue requires a separate study and will be a topic of a forthcoming work.

\section{Electron-phonon interaction}

Now we demonstrate that the electron-phonon interaction leads to the electron evolution which satisfies the conditions of the quantum $H$-theorem. In a standard electron-phonon interaction model, a representative test electron `sees' a screened short-range ionic potential. Since the resulting scattering time is short and the ion is much heavier than the electron, the position of a given ion remains nearly unchanged during the interaction with the electron, see Fig.~\ref{fig:lattice}. Then the standard consideration of the electron-phonon interaction\cite{Pitaevskii:1981} results in a conclusion that at high temperatures where the typical energy of an electron is relatively high (i.e. it exceeds the Debye energy, the maximal energy of phonons), the electron-phonon collisions are quasi-elastic. This allows us to apply the same arguments as for the model of an electron interacting with two-level impurities considered above. Indeed, the slow ion-dynamics preserves the classical distribution function, $\rho_\mathrm{ion}(\{\mathbf{r}\})$, for the ions positions $\{\mathbf{r}\} = \{\mathbf{r}_1,\,\mathbf{r}_2, \ldots \} $ and hence the induced ${\hat F}$-operators are diagonal in the ion coordinate basis,
\begin{equation}
	{\hat F}_{ji} 
	= \int \prod_{\alpha}d\mathbf{r}_\alpha\, s_{ji}(\{\mathbf{r}\}) \rho_\mathrm{ion}(\{\mathbf{r}\}) \, \prod_{\alpha}|\mathbf{r}_\alpha\rangle \langle \mathbf{r}_\alpha|,
	\label{eq:Fphonon}
\end{equation}
where $s_{ji}(\{\mathbf{r}\})$ are electron scattering matrix on the potential defined by the ions positions $\{\mathbf{r}\}$. Therefore the conditions of the quantum $H$-theorem hold for electrons with energies exceeding the Debye energy. The initial disentanglement of the electron from the ionic reservoir is ensured either by the fact that a given ion did not participate in the previous collisions with the electron, or has already `forgotten' about such an event due to the monogamy of entanglement.

It is noteworthy that unlike the other systems discussed in previous sections, the evolution of the electron-phonon system does not occur in the unital channel. Nevertheless, it satisfies the condition of the $H$-theorem thus illustrating a mathematical result establishing that for some classes of initial states even the evolution defined by non-unital channels may occur with the non-diminishing entropy.\cite{Amosov:2015}

\subsection*{Acknowledgements}

We are delighted to thank G. Blatter, G.-M. Graf, M. McBreen, L. B. Ioffe and G. G. Amosov for illuminating discussions. The work was supported by the RFBR Grant No. 14-02-01287 (G.B.L.), by the Pauli Center for Theoretical Studies at ETH Zurich (G.B.L.), by the US Department of Energy, Office of Science, Materials Sciences and Engineering Division (I.A.S. and V.M.V.), and by the Swiss National Foundation through the NCCR QSIT (A.V.L.).

\appendix

\section{Unitarity constraints} \label{sec:unitarity_constraints}

Consider a quantum system (particle) which interacts with a quantum reservoir. The evolution of the joint grand system (particle plus reservoir) can be presented in the form,
\begin{equation}
	|\psi_i\rangle |n\rangle \to \sum_{jm} U_{jm,in} \, |\psi_j\rangle |m\rangle,
	\label{eq:GSS1}
\end{equation}
where $|\psi_i\rangle$ and $|n\rangle$ are some orthonormal complete sets of states in the particle and reservoir Hilbert space respectively. The transition amplitudes $U_{jm,in}$ comprise a unitary matrix ${\hat U}$: ${\hat U}^\dag {\hat U} = {\hat U} {\hat U}^\dag = \I$.

We assume no energy exchange between the quantum system and the reservoir. Let us consider a fixed energy subspace $E$ of the particle Hilbert space spanned by a basis $|\psi_{i,\E}\rangle$, here index $i$ denotes all remaining degrees of freedom of the particle. Then one can represent the grand evolution operator ${\hat U}$ in the form,
\begin{equation}
	{\hat U} 
	= \sum_E |\psi_{j,\E}\rangle \langle \psi_{i,\E}|\, s_{ji,\E} {\hat F}_{ji,\E},
\end{equation}
Then the unitarity constraints can be rewritten in the form,
\begin{align}
	&\sum_i s_{ij',\E}^* s_{ij,\E} \, \langle m^\prime | {\hat F}_{ij',\E}^\dag {\hat F}_{ij,\E}| m\rangle = \delta_{jj'}\delta_{mm'},
	\label{eq:GUNTR3}
	\\
	&\sum_i s_{ji,\E} s_{j^\prime i,\E}^*\, \langle m| {\hat F}_{ji,\E} {\hat F}_{j^\prime i,\E}^\dag| m^\prime \rangle = \delta_{jj^\prime} \delta_{mm^\prime}.
\end{align}

\section{Proof of $H$-theorem} \label{sec:theorem_proof}

Here we derive Eq.~\eqref{eq:PHII} and sketch the proof of the $H$-theorem. We consider a quantum system which interacts with a quantum reservoir. The evolution operator ${\hat U}$ is unitary, ${\hat U} {\hat U}^\dag = 1$, and using Eq.~\eqref{eq:QES} we write the unitarity constraints in the form,
\begin{equation}
	\sum_i s_{ji,\E} s_{j'i,\E}^*\, \langle m| {\hat F}_{ji,\E} {\hat F}_{j'i,\E}^\dag| m' \rangle = \delta_{jj'} \delta_{mm'}.
	\label{eq:GUNTR4}
\end{equation}

Let the reservoir initially be in a state described by the density matrix $\hat\pi = \sum_{nn'} \pi_{nn'} \, |n\rangle \langle n'|$. Then
\begin{equation}
	[\Phi(\hat\rho)]_{jj'} =
	\sum_{ii'} \rho_{ii'} \tr \{ \langle \psi_{j}| {\hat U} |\psi_i\rangle \, \hat\pi\, \langle \psi_{i'}| {\hat U}^\dag|\psi_{j'}\rangle \}
	\label{eq:QCH1}
\end{equation}
where $\hat\rho = \sum_{ii'} \rho_{ii'} \, |\psi_i\rangle \langle \psi_{i'}|$.

The system states at different energies are transformed independently. Let us consider a system density matrix at a fixed energy subspace of the particle Hilbert space, ${\hat\rho}_\E = \sum_{ij} \rho_{ij,\E} \, |\psi_{i,\E}\rangle \langle \psi_{j,\E}|$. Then making use of the factorization representation of the evolution operator, see Eq.~\eqref{eq:QCH1}, one can rewrite the quantum channel in the form,
\begin{equation}
	[\Phi({\hat\rho}_\E)]_{jj'} =
	\sum_{ii'} \rho_{ii',\E} \, s_{ji,\E} s_{j'i',\E}^* \, \langle {\hat F}_{j'i',\E}^\dag {\hat F}_{ji,\E}\rangle,
	\label{eq:QCH2}
\end{equation}
where the averaging of the reservoir operators ${\hat F}_{ji,\E}$ is done with respect to the density matrix $\hat\pi$: $\langle \dots \rangle = \tr \{\hat\pi \dots\}$. Now let us consider transformation of $\rho_{ii',\E} = \delta_{ii'} = \I_\E$. Then using Eq.~\eqref{eq:QCH2} we find:

\begin{equation}
	[\Phi(\I_\E)]_{jj'} = 
	\sum_i s_{ji,\E}s_{j'i,\E}^*\, \langle {\hat F}_{j'i,\E}^\dag {\hat F}_{ji,\E}\rangle.
	\label{eq:UNTL2}
\end{equation}
Averaging Eq.~\eqref{eq:GUNTR4} with respect to the density matrix of the system, moving the left hand side of the averaged equation to the right, and adding the result to the rhs of Eq.~\eqref{eq:UNTL2}, one arrives at Eq.~\eqref{eq:PHII} of the text:
\begin{equation}
	[\Phi(\I_\E)]_{jj'} =
	\delta_{jj'} 
	+ \sum_i s_{ji,\E}s_{j'i,\E}^*\, \langle [{\hat F}_{j'i,\E}^\dag, {\hat F}_{ji,\E}] \rangle.
	\label{eq:UNTL3}
\end{equation}

Now we outline the proof of the $H$-theorem using Eq.~\eqref{eq:entropy_gain}. To that end we have to calculate $\Phi(\I) = \Phi(\prod_\E \otimes \I_\E)$. Let $[\Phi(\I_\E)]_{jj'}= \delta_{jj'}$ on some subset $\mathbf{\cal E} = \{E\}$. Then using Eq.~\eqref{eq:UNTL2} in the rhs of Eq.~\eqref{eq:entropy_gain}, we obtain
\begin{equation}
	-k_\B\sum_{jj',\E} \langle \psi_{j,\E}|\Phi({\hat\rho})|\psi_{j',\E}\rangle\langle \psi_{j',\E}|\log\Phi(\I) |\psi_{j,\E}\rangle.
	\label{eq:trace}
\end{equation}
Accordingly, in Eq.~\eqref{eq:trace} the terms for which $E \in \mathbf{\cal E}$, become
\begin{equation}
	\langle\psi_{j,\E}| \Phi({\hat\rho}) |\psi_{j,\E}\rangle \langle \psi_{j,\E}| \log 1 |\psi_{j,\E}\rangle = 0.
	\nonumber
\end{equation}
Now if $ \langle \psi_{i,\E}|\,{\hat\rho}\,|\psi_{i,\E}\rangle\neq 0$ only for $E \in \mathbf{\cal E}$ (therefore, for energies $E\notin \mathbf{\cal E}$, all matrix elements $ \langle \psi_{i,\E}|\,{\hat\rho}\,|\psi_{j,\E}\rangle= 0$), the entire expression~\eqref{eq:trace} is zero, Eq.~\eqref{eq:entropy_gain} yields $\Delta S \geqslant 0$, and the theorem is proved.

\section{Example of a non-unital channel} \label{sec:non_unital_channel}

Let us inspect a charged particle moving in a three-lead conductor and interacting with the spin via induced magnetic field, see Fig.~\ref{fig:Y-junction} in the main text, and, according to the framework of our general consideration, initially disentangled from the spin. In the absence of the external magnetic field the energy exchange is absent and the particle is energy-isolated. The joint scattering states of the particle and spin have the form
\begin{equation}
	|\psi_\alpha^{\scriptscriptstyle(\mathrm{in})}\rangle |\sigma_0\rangle 
	\to \sum_\beta s_{\beta\alpha}\, |\psi_\beta^{\scriptscriptstyle(\mathrm{out})}\rangle \, 
	{\hat U}_\beta {\hat U}_\alpha^\dag \, |\sigma_0\rangle,
\end{equation}
where $|\psi_\alpha^{\scriptscriptstyle(\mathrm{in/out})}\rangle$ is the particle's incoming/outgoing state in the lead $\alpha$, $s_{\beta\alpha}$ are the components of the scattering matrix of the three lead set up, $|\sigma_0\rangle$ is the initial state of the spin and ${\hat U}_\alpha$ (${\hat U}_\alpha^\dag$) is the unitary spin-$1/2$ rotation of the spin due to outgoing (incoming) electron in the lead $\alpha$. Then the operators ${\hat F}$ in Eq.~\eqref{eq:QES} of the main text are defined as ${\hat F}_{\beta\alpha}={\hat U}_\beta {\hat U}_\alpha^\dag$.

We choose spin-$1/2$ rotations as ${\hat U}_1 = \I$, ${\hat U}_2 = i\hat\sigma_x$ and ${\hat U}_3 = i\hat\sigma_y$, where $\hat\sigma_x$ and $\hat\sigma_y$ are the Pauli matrices. In this case $\bigl[{\hat U}_\alpha, {\hat U}_\beta\bigr] \neq 0$. Accordingly, $\langle [{\hat F}_{\beta^\prime \alpha}^\dag, {\hat F}_{\beta\alpha}] \rangle\neq 0$, and the resulting quantum channel is non-unital. Then,
\begin{align}
	\Phi(\I)
	= \I
	& - \bigl\{ 2i \, |1\rangle \langle 2| \, (s_{11}s_{21}^*+s_{12}s_{22}^*) \, \langle\hat\sigma_x\rangle + \mathrm{H.c.} \bigr\}
	\nonumber
	\\
	& - \bigl\{ 2i \, |1\rangle \langle 3| \, (s_{11}s_{31}^* + s_{13}s_{33}^*) \, \langle\hat\sigma_y\rangle + \mathrm{H.c.} \bigr\}
	\nonumber
	\\
	& + \bigl\{ 2i \, |2\rangle \langle3| \, (s_{22}s_{32}^* + s_{23}s_{33}^*) \, \langle\hat\sigma_z\rangle + \mathrm{H.c.} \bigr\}.
\end{align}
To simplify this equation, we assume time-reversal symmetry of the electron scattering, $s_{ij} = s_{ji}$, and exploiting the following representation\cite{Lesovik:2001} of the elastic scattering matrix,
\begin{equation}
	s_{ii}
	= \frac12 s_{jk}^* s_{ij}s_{ik} \Bigl( \frac1{T_{jk}}
	- \frac1{T_{ij}} - \frac1{T_{ik}}\Bigr),
	\quad i \neq j \neq k,
\end{equation}
where $T_{ij} = |s_{ij}|^2$ are transmission probabilities, we obtain Eq.~\eqref{eq:PHII3} of the main text.

\section{One-dimensional scattering} \label{sec:1d_scattering}

From the unitarity of the elastic scattering matrix it follows $|s_\RR|^2 + |s_\LR|^2 = 1$, $|s_\RL|^2 + |s_\LL|^2 = 1$, $s_\LL^* s_\LR^{\phantom *} + s_\RL^*s_\RR^{\phantom *} = 0$, and $s_\RL^{\phantom *} s_\LL^* + s_\LR^* s_\RR^{\phantom *} = 0$. The unitarity of the joint particle-reservoir gives us
\begin{equation}
	s_\LL^* s_\LR^{\phantom *} \, \langle n''|n\rangle
	+ s_\RL^*s_\RR^{\phantom *} \, \langle n|n'\rangle
	= 0.
	\label{eq:1Dunitarity}
\end{equation}
Hence $\langle n''|n\rangle = \langle n|n'\rangle$ and
\begin{equation}
	s_\LL^*s_\RL^{\phantom *} \langle n''|n\rangle
	+ s_\LR^* s_\RR^{\phantom *} \langle n|n'\rangle
	= 0.
	\label{eq:1Dunitarity2}
\end{equation}

Let us consider the transformation of the identity operator $\I = \cR \langle\mathrm{R}| +\cL \bL$ if reservoir initially rests in a state $|n\rangle \langle n|$,
\begin{align}
	\I
	\to \, & \cR \langle\mathrm{R}|\, (|s_\RL|^2 + |s_\RR|^2)
	+ \cL \bL\, (|s_\LL|^2+|s_\LR|^2) \, +
	\nonumber \\
	& \bigl[\cR \bL\,(s_\LL^*s_\RL^{\phantom *} \langle n''|n\rangle
	+ s_\LR^* s_\RR^{\phantom *} \langle n|n'\rangle) + \mathrm{H.c.} \bigr].
\end{align}
One can see that the off-diagonal terms $\cR \bL$ of the transformed identity operator vanish due to unitarity constraint, see Eq.~\eqref{eq:1Dunitarity2}, while the diagonal elements sum to the identity operator and hence $\Phi(\I) = \I$. The same arguments are valid if initially reservoir rests in an arbitrary mixed state non-entangled with the particle.

The same arguments can be given in a more general form, exploiting the Theorem of the main text. Making use of the scattering states [Eqs.~\eqref{eq:scattering_state1}--\eqref{eq:scattering_state2} of the main text] one can construct the joint evolution operator at a given energy of the incoming particle,
\begin{align*}
	{\hat U}
	& = s_\LR \cL \bL \otimes \I
	+ s_\RL \cR \bR \otimes \I \\
	& + s_\RR \cR \bL \otimes {\hat F}_\RR
	+ s_\LL \cL \bR \otimes {\hat F}_\LL,
\end{align*}
which transforms the incoming particle states into outgoing states. Here, first two terms corresponds to the forward scattering of the particle which does not change the reservoir state. The last two terms accounts for the backscattering events which do change the reservoir state. The joint unitarity constraint ${\hat U}^\dag {\hat U} = \I$ gives,
\begin{align}
	& \bigl( |s_\LR|^2 \I + |s_\RR|^2 {\hat F}_\RR^\dag {\hat F}_\RR^{\phantom\dag} \bigr) \cL \bL + \nonumber \\
	& \bigl( |s_\RL|^2 \I + |s_\LL|^2 {\hat F}_\LL^\dag {\hat F}_\RR^{\phantom\dag} \bigr) \cR \langle\mathrm{R}| + \nonumber \\
	& \bigr[ \bigl( s_\LL^* s_\LR^{\phantom *} {\hat F}_\LL^\dag + s_\RL^* s_\RR^{\phantom *} {\hat F}_\RR\bigr) \cR \bL + \mathrm{H.c.} \bigr] = \I.
	\label{eq:UU1}
\end{align}
The unitarity of the elastic scattering amplitudes $s_\RL^* s_\RR + s_\LL^* s_\LR = 0$ vanish the off-diagonal terms $\cL \langle\mathrm{R}|$ and $\cR \bL$ on the Eq.~\eqref{eq:UU1} if
\begin{equation}
	{\hat F}_\RR^{\phantom\dag} = {\hat F}_\LL^\dag,
\end{equation}
while the normalization condition $|s_\LR|^2 + |s_\RR|^2 = 1$ and $|s_\RL|^2 +|s_\LL|^2 = 1$ gives,
\begin{equation}
	{\hat F}_\RR^\dag {\hat F}_\RR^{\phantom\dag} = \I, \quad
	{\hat F}_\LL^\dag {\hat F}_\LL^{\phantom\dag} = \I
\end{equation}
and hence ${\hat F}_\RR^{\phantom\dag}$ and ${\hat F}_\LL^{\phantom\dag} = {\hat F}_\RR^\dag$ are unitary operators. It follows then that $[{\hat F}_{\beta' \alpha}, {\hat F}_{\beta\alpha}] = 0$ for $\alpha$, $\beta$, $\beta' \in \{\mathrm{L}, \mathrm{R}\}$ and the scattering process [Eqs.~\eqref{eq:scattering_state1}--\eqref{eq:scattering_state2} of the main text] describes an unital quantum channel.

\end{document}